\documentclass[twocolumn,showpacs,aps,prl,letterpaper]{revtex4}

\usepackage{amsmath}
\usepackage{amssymb}
\usepackage{multirow}
\usepackage{graphicx}
\usepackage{float}

\RequirePackage{xspace}
\allowdisplaybreaks

\begin{document}

\title{\large \bfseries \boldmath Study of $CP$ violation in $D\to VV$ decay at BES-III}
\author{Xian-Wei Kang$^{1,2}$}\email{kangxw@ihep.ac.cn}
\author{Hai-Bo Li$^1$}\email{lihb@ihep.ac.cn}
\affiliation{$^1$Institute of High Energy Physics, P.O.Box 918,
Beijing 100049, China\\ $^2$Department of Physics, Henan Normal
University, Xinxiang 453007, China}

\begin{abstract}

In this paper, we intend to study the problem of $CP$ violation in
$D$ meson by $D\to VV$ decay mode in which the T violating
triple-product correlation is examined. That would undoubtedly be
another excellent probe of New Physics beyond Standard Model. For
the neutral $D$, we focus on direct $CP$ violation without
considering $D^0-\bar D^0$ oscillation. Experimentally, by a full
angular analysis one may obtain such $CP$ violating signals, and
particularly it is worth mentioning that the upcoming large $D$
data samples at BES-III in Beijing will provide a great
opportunity to perform it.

\end{abstract}


\maketitle


The topic of $CP$ violation in the $D$-meson sector has been the
subject of extensive studies involving both charged and neutral D
meson decays these years
\cite{E791,Buccela,Aubert,Boca,Cronin,Nandy,zz,Oliver,Stephen,Gordon,Monich,Korner1}.
In a recent reference~\cite{Jerome}, they exploited the angular
and quantum correlation in the $D^0\bar{D}^0$ pairs produced
through the decay of the $\psi(3770)$ resonance at BES-III to
investigate $CP$ violation. They build $CP$ violating observables
in $e^+e^-\rightarrow \psi(3770)\rightarrow
D^0\bar{D}^0\rightarrow (V_1V_2)(V_3 V_4)$ ($V$ denotes vector
meson) to isolate specific New Physics effects in the charm
sector~\cite{Jerome}.

In practice, one can also probe the $CP$ violation in $D$ meson by
one $D$ decay without considering the quantum correlation, which
is the motivation of this paper. Among the various kinds of $D$
decay modes, $D\to V_1V_2 $ and subsequently decaying to two
pseudoscalars for each vector meson is a particularly interesting
one in the perspective of the copious kinematics of final state
interaction (FSI). As the same case in $B$
meson~\cite{Valencia,A.Datta}, the new type of $T$ violating
signal involving so-called triple-product (TP) will emerge by
comparing a pair of $CP$ conjugate processes, where TP is composed
of the momentum of one vector meson and two polarizations.
Assuming $CPT$ invariance, $T$ violating TP asymmetry is
equivalent to $CP$ violating. We shall see in this letter such TP
asymmetries are related to the helicity components in the angular
distribution of the $D\to V_1V_2$ process.  Moreover, performing a
full angular analysis is feasible and realistic in experiment.

On the other hand, TP asymmetry (in the latter section we
sometimes also refer to TP asymmetry as $CP$ asymmetry assuming
$CPT$ invariance) is a sensitive signal of New Physics. As is well
known that Standard Model (SM) predictions for $CP$ violation in
the charm sector are very small,
thus any significant such signals would be exciting. Currently,
the BES-III experiment is collecting data at $\psi(3770)$ peak. In
the short future, lots of events of $D$ decay will be accumulated,
which will provide a great opportunity to perform a full angular
analysis to further achieve a valuable information for the
question discussed here.

Let us first consider the process $D(p)\to
V_1(k,\epsilon_1)V_2(q,\epsilon_2)$, where the two vectors
$V_1,V_2$ are characterized as their four-momenta and
polarizations $(k,\epsilon_1)$ and $(q,\epsilon_2)$, respectively.
We can write the most general invariant amplitude as a sum of
three terms that we will call $s$, $d$, $p$
~\cite{Valencia,A.Datta,Kramer,Tseng},
\begin{eqnarray}\label{M}
\mathcal{M}&\equiv& as + bd + icp \\\nonumber
 &=& a \epsilon_1^*\cdot\epsilon_2^* + \frac{b}{m_1m_2}
 (p\cdot\epsilon_1^*)(p\cdot\epsilon_2^*)\\ \nonumber
 &&\qquad + i
 \frac{c}{m_1m_2}\epsilon^{\alpha\beta\gamma\delta}\epsilon_{1\alpha}^*\epsilon_{2\beta}^*k_{\gamma}p_{\delta}\,,
\end{eqnarray}
where $m_1$ ($m_2$) is the mass of $V_1$ ($V_2$), and the scalar
coefficients $a$, $b$ and $c$ are generally complex and can
receive contributions from several amplitudes with different
phases. Thus, one can parameterize the coefficients as
\cite{Valencia}
\begin{eqnarray}\label{abc}
a &=& \sum_j a_j e^{i\delta_{sj}}e^{i\phi_{sj}}\,,\\ \nonumber
 b &=&\sum_j b_j e^{i\delta_{dj}}e^{i\phi_{dj}}\,,\\ \nonumber
c &=&\sum_j c_j e^{i\delta_{pj}}e^{i\phi_{pj}}\,,
\end{eqnarray}
where $a_j$, $b_j$ and $c_j$ are the modulus of their
corresponding complex quantities, and $\delta_j$ denotes strong
phase (also called unitary phase in Ref.~\cite{Valencia}), and
$\phi_j$ is the weak phase which is the necessary condition for
occurring of $CP$ violation on the basis of
Cabibbo-Kobayashi-Maskawa (CKM) mechanism~\cite{Cabibbo,Maskawa}
in the SM. Squaring the matrix element of Eq.~\eqref{M}, one
obtains
\begin{widetext}
\begin{eqnarray}\label{M square}
|\mathcal{M}|^2& = &|a|^2|\epsilon_1^*\cdot\epsilon_2^*|^2 +
\frac{|b|^2}{m_1^2m_2^2}|(k\cdot\epsilon_2^*)(q\cdot\epsilon_1^*)|^2
+
\frac{|c|^2}{m_1^2m_2^2}|\epsilon^{\alpha\beta\gamma\delta}\epsilon_{1\alpha}^*\epsilon_{2\beta}^*k_{\gamma}p_{\delta}|^2
+2\frac{Re(ab^*)}{m_1m_2}(\epsilon_1^*\cdot\epsilon_2^*)(k\cdot\epsilon_2^*)(q\cdot\epsilon_1^*)\\
\nonumber
&&+2\frac{Im(ac^*)}{m_1m_2}(\epsilon_1^*\cdot\epsilon_2^*)
\epsilon^{\alpha\beta\gamma\delta}\epsilon_{1\alpha}^*\epsilon_{2\beta}^*k_{\gamma}p_{\delta}+2\frac{Im(bc^*)}{m_1^2m_2^2}
(k\cdot\epsilon_2^*)(q\cdot\epsilon_1^*)\epsilon^{\alpha\beta\gamma\delta}\epsilon_{1\alpha}^*\epsilon_{2\beta}^*k_{\gamma}p_{\delta}\,.
\end{eqnarray}
\end{widetext}

Next, using $CPT$ invariance, the matrix element for the
antiparticle decay $\bar D(p)\to \bar{V}_1(k,\epsilon_1)\bar
V_2(q,\epsilon_2)$ can be written as
\begin{eqnarray}\label{bar M}
 \overline{\mathcal{M}}& = &\bar a \epsilon_1^*\cdot\epsilon_2^* + \frac{\bar
 b}{m_1m_2}(p\cdot\epsilon_1^*)(p\cdot\epsilon_2^*)\\ \nonumber
 && -i \frac{\bar
 c}{m_1m_2}\epsilon^{\alpha\beta\gamma\delta}\epsilon_{1\alpha}^*\epsilon_{2\beta}^*k_{\gamma}p_{\delta}\,,
\end{eqnarray}
with
\begin{eqnarray}\label{bar abc}
\bar a &=& \sum_j a_j e^{i\delta_{sj}}e^{-i\phi_{sj}}\,,\\
\nonumber
 \bar b &=&\sum_j b_j e^{i\delta_{dj}}e^{-i\phi_{dj}}\,,\\ \nonumber
\bar c &=&\sum_j c_j e^{i\delta_{pj}}e^{-i\phi_{pj}}\,.
\end{eqnarray}
Note that $CP$ operator leaves strong phases invariant and only
changes the sign of weak phase. From Eq.~\eqref{M} and
Eq.~\eqref{bar M}, we will find that the $p$ wave amplitude in
$\bar{\mathcal{M}}$ changes the sign comparing with $\mathcal{M}$,
which will induce an interesting property between
$|\bar{\mathcal{M}}|^2$ and $|\mathcal{M}|^2$.  To be clear, we
would square the matrix element for the antiparticle decay in
Eq.~\eqref{bar M}:
\begin{widetext}
\begin{eqnarray}\label{bar M square}
|\overline{\mathcal{M}}|^2& = &|\bar
a|^2|\epsilon_1^*\cdot\epsilon_2^*|^2 + \frac{|\bar
b|^2}{m_1^2m_2^2}|(k\cdot\epsilon_2^*)(q\cdot\epsilon_1^*)|^2 +
\frac{|\bar
c|^2}{m_1^2m_2^2}|\epsilon^{\alpha\beta\gamma\delta}\epsilon_{1\alpha}^*\epsilon_{2\beta}^*k_{\gamma}p_{\delta}|^2
+2\frac{Re(\bar a\bar b^*)}{m_1m_2}(\epsilon_1^*\cdot\epsilon_2^*)(k\cdot\epsilon_2^*)(q\cdot\epsilon_1^*)\\
\nonumber &&-2\frac{Im(\bar a\bar
c^*)}{m_1m_2}(\epsilon_1^*\cdot\epsilon_2^*)
\epsilon^{\alpha\beta\gamma\delta}\epsilon_{1\alpha}^*\epsilon_{2\beta}^*k_{\gamma}p_{\delta}-2\frac{Im(\bar
b\bar c^*)}{m_1^2m_2^2}
(k\cdot\epsilon_2^*)(q\cdot\epsilon_1^*)\epsilon^{\alpha\beta\gamma\delta}\epsilon_{1\alpha}^*\epsilon_{2\beta}^*k_{\gamma}p_{\delta}\,.
\end{eqnarray}

For $D^0\to V_1V_2$ decay, one can define an asymmetry
$\mathcal{A_T}$ with the definite sign for the triple product
$(\vec{k}\cdot\vec{\epsilon_1^*}\times\vec{\epsilon_2^*})$
as~\cite{Valencia}
\begin{equation}\label{asymmetry_N}
\mathcal{A_T} =
\frac{N(\vec{k}\cdot\vec{\epsilon_1^*}\times\vec{\epsilon_2^*}>0)-N(\vec{k}\cdot\vec{\epsilon_1^*}\times\vec{\epsilon_2^*}<0)}
{N_{total}}\,,
\end{equation}
where the subscript $\mathcal{T}$ implies triple products and $N$
denotes the corresponding number of events. Eq.~\eqref{asymmetry_N}
above is actually
\begin{equation}\label{asymmetry_Gamma}
\mathcal{A_T} =
\frac{\Gamma(\vec{k}\cdot\vec{\epsilon_1^*}\times\vec{\epsilon_2^*}>0)-\Gamma(\vec{k}\cdot\vec{\epsilon_1^*}\times\vec{\epsilon_2^*}<0)}
{\Gamma(\vec{k}\cdot\vec{\epsilon_1^*}\times\vec{\epsilon_2^*}>0)+\Gamma(\vec{k}\cdot\vec{\epsilon_1^*}\times\vec{\epsilon_2^*}>0}\,.
\end{equation}
Similarly, for $\bar D^0\to \bar V_1\bar V_2$ decay,
$\bar{\mathcal{A_T}}$ can also be constructed as the same way. In
$|\mathcal{M}|^2$, a triple-product correlation arises from
interference terms involving the $p$ amplitude, and will be
present if $Im(ac^*)$ \big(or $Im(bc*)$\big) is non-zero. After a
simple calculation by inserting Eq.~\eqref{abc} and Eq.~\eqref{bar
abc}, we see that

\begin{equation}
\mathcal{A_T}\varpropto Im(ac^*) =
\sum_{i,j}a_ic_j\sin[(\phi_{si}-\phi_{pj})+(\delta_{si} -
\delta_{pj})]
\end{equation}
Note that a non-zero triple correlation does not necessarily imply
$CP$ violation, since final state interactions (FSI) can fake it,
namely the strong phase can also produce non-zero $\mathcal{A_T}$
(or $\bar{\mathcal{A_T}}$)  even the weak phases are zero. Yet
comparing a triple correlation for $CP$ conjugate transitions
allows to distinguish genuine $CP$ violation from FSI effects.
Thus we obtain,
\begin{eqnarray}\label{TP_asymmetry1}
\frac{1}{2}(\mathcal{A_T} + \bar{\mathcal{A_T}})\varpropto
\frac{1}{2}[Im(ac^*)- Im(\bar a \bar c^*)]
&=&\sum_{i,j}a_ic_j\sin(\phi_{si}-\phi_{pj})\cos(\delta_{si} -
\delta_{pj})\,, \end{eqnarray} and
\begin{eqnarray}\label{TP_asymmetry2} \frac{1}{2}(\mathcal{A_T} -
\bar{\mathcal{A_T}})\varpropto\frac{1}{2}[Im(ac^*) + Im(\bar a
\bar c^*)] &=& \sum_{i,j}a_ic_j\cos(\phi_{si} -
\phi_{pj})\sin(\delta_{si} - \delta_{pj})\,.
\end{eqnarray}
So far, a non-zero value $\mathcal{A_T} + \bar{\mathcal{A_T}}$
will be undoubtedly a clean signal of $CP$ non-conservation,
because there must be at least one non-zero weak phase $\phi$.
Here, we also note that if there is only one amplitude
contributing to each partial wave, one can simultaneously
determine the strong phase difference and weak phase difference
from Eq.~\eqref{TP_asymmetry1} and Eq.~\eqref{TP_asymmetry2}.
\end{widetext}

 Experimentally, one can perform a full angular analysis to obtain
 the above TP asymmetry information because the complex coefficients
 $a$, $b$, $c$ are related to the helicity amplitudes $A_0$,
 $A_{||}$, $A_{\perp}$ as discussed in Refs.~\cite{Kramer,Rosner},
 \begin{eqnarray}\label{connection}
A_{0} &=& -ax-b(x^2-1),\nonumber\\ A_{||} &=&
\sqrt{2}a,\nonumber\\A_{\perp}& =& \sqrt{2}c\sqrt{x^2-1}\,,
 \end{eqnarray}
 where
 \begin{equation}
 x = \frac{k\cdot q}{m_1m_2} =
 \frac{m^2_D-m_1^2-m_2^2}{2m_1m_2}\,,
 \end{equation}
where $m_D$ is the mass of $D$ meson.

 Now we turn to the full angular dependence of process $D\to
 V_1V_2\to (P_1P_2)(P_3P_4)$ with $P$ pseudoscalar, after some algebra one can get

\begin{figure}[htbp]
\centering
\includegraphics[scale=0.5]{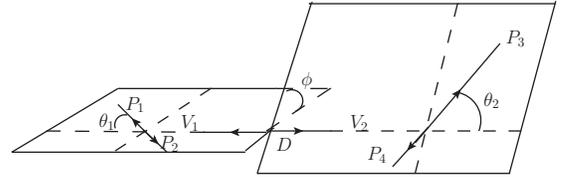}
\caption{Illustrative plot for the decay kinematics of process $D\to
V_1V_2\to (P_1P_2) (P_3P_4)$ in the rest frame of $V_{1,2}$.}
\label{Fig}
\end{figure}

 \begin{widetext}

 \begin{eqnarray} \label{angular}
 \frac{d\Gamma}{d\cos\theta_1d\cos\theta_2d\phi}\varpropto
 \frac{1}{2}\sin^2\theta_1\sin^2\theta_2\cos^2\phi|A_{||}|^2 +
 \frac{1}{2}\sin^2\theta_1\sin^2\theta_2\sin^2\phi|A_{\perp}|^2 +
 \cos^2\theta_1\cos^2\theta_2|A_0|^2 \nonumber \\
 - \frac{1}{2}\sin^2\theta_1\sin^2\theta_2\sin2\phi
 Im(A_{\perp}A_{||}^*)-\frac{\sqrt{2}}{4}\sin 2\theta_1\sin
 2\theta_2 \cos\phi Re(A_{||}A_0^*) +
 \frac{\sqrt{2}}{4}\sin2\theta_1\sin2\theta_2\sin\phi
 Im(A_{\perp}A_0^*)\,,
 \end{eqnarray}
\end{widetext}
where we have introduced $A_{\perp}$ with definite odd $CP$
eigenvalue and the $CP$ even partners $A_{0}$, $A_{||}$ via
\begin{eqnarray}
A_0& = &A_0\,,\nonumber \\
A_{||}& = & \frac{1}{\sqrt{2}} (A_{11} + A_{-1-1})\,,\nonumber \\
A_{\perp}& =& \frac{1}{\sqrt{2}} (A_{11} - A_{-1-1})\,,
\end{eqnarray}
with $A_{\lambda_1\lambda_2}$ denoting the helicity mode of two
vector mesons. $\theta_i's$ (i=1,2) are the angles between the
direction of motion of one of the $V_{1,2}\to PP$ pseudoscalar
final states and the inverse direction of motion of the $D$ meson
as measured in the $V_{1,2}$ rest frame, $\phi$ is the angle
between the two decay plane of vector mesons in the $D$ rest
frame. Figure~\ref{Fig} illustrates the decay kinematics of the
process $D\to V_1V_2\to (P_1P_2) (P_3P_4)$ in the rest frame of
$V_{1,2}$. Eq.~\eqref{angular} is consistent with that from Refs.~
\cite{Korner2,Deshan}.

As discussed previously, the TP asymmetry is connected with the
angular dependence, combining Eqs.~\eqref{TP_asymmetry1},
\eqref{TP_asymmetry2}, \eqref{connection}, one can define the
following $T$-odd quantities~\cite{Valencia,A.Datta},
\begin{equation}
\mathcal{A_T}^0 \equiv \frac{Im(A_{\perp}A_0^*)}{|A_0|^2 +
|A_{\perp}|^2 + |A_{||}|^2}\,,
\end{equation}
and
\begin{equation}
\mathcal{A_T^{||}} \equiv \frac{Im(A_{\perp}A_{||}^*)}{|A_0|^2 +
|A_{\perp}|^2 + |A_{||}|^2},
\end{equation}
thus we will derive the $CP$ violating observables,
\begin{eqnarray}
\mathcal{A}& =&  \frac{1}{2}(\mathcal{A_T}^0 + \bar{
\mathcal{A_T}}^0)\\ \nonumber & =&
\frac{1}{2}\Big(\frac{Im(A_{\perp}A_0^*)}{|A_0|^2+|A_{\perp}|^2+|A_{||}|^2}
+ \frac{Im(\bar {A}_{\perp}\bar{A}_0^*)}{|\bar {A}_0|^2+|\bar
{A}_{\perp}|^2+|\bar {A}_{||}|^2}\Big)\,,
\end{eqnarray}
and
\begin{eqnarray}
 \mathcal{A}'&=& \frac{1}{2}(\mathcal{A_T}^{||} + \bar
 {\mathcal{A_T}}^{||})\\ \nonumber
&=&
\frac{1}{2}\Big(\frac{Im(A_{\perp}A_{||}^*)}{|A_0|^2+|A_{\perp}|^2+|A_{||}|^2}
+ \frac{Im(\bar {A}_{\perp}\bar {A}_{||}^*)}{|\bar {A}_0|^2+|\bar
{A}_{\perp}|^2+|\bar {A}_{||}|^2}\Big)\,.
\end{eqnarray}

Before this study, there had been attempts to study $CP$ violation
of $D$ meson via $T$ violating TP correlation in theoretical
viewpoint that differs from our method~\cite{Korner1,Bigi1,Bigi2}.
But the only reported experimental search for $T$-odd asymmetries
is from FOCUS in the $D^0\to K^+K^-\pi^+\pi^-$ and $D^+_S\to K_S
K^+\pi^+\pi^-$ decay modes~\cite{FOCUS}, as listed in Table
\ref{tab2}. No evidence for a T asymmetry is observed. The large
BES-III data sample is expected to provide enhanced sensitivity to
possible $T$ violating asymmetries.

\begin{table}[H]
\begin{center}
\begin{tabular}{cc}\hline\hline
  Decay mode   &$\mathcal{A}$(\%) \\
\hline
  $D^0\to K^+K^-\pi^+\pi^-$       &$1.0\pm5.7\pm3.7$ \\
  $D^+\to K_S K^+\pi^+\pi^-$  &$2.3\pm6.2\pm2.2$  \\
  $D_S^+\to K_S K^+\pi^+\pi^-$       &$-3.6\pm6.7\pm2.3$ \\
\hline\hline
\end{tabular}
\caption{$T$ violating asymmetries in $D$ meson decays from the
FOCUS experiment~\cite{FOCUS}.}\label{tab2}
\end{center}
\end{table}

At last, we consider the potential sensitivity on the $CP$
violating observables $\mathcal{A}$ and $\mathcal{A'}$ at BES-III.
From Eq.~\eqref{asymmetry_N}, for a small asymmetry, there is a
general result that it's error is approximately estimated as
$1/\sqrt{N_{total}}$, where $N_{total}$ is the total number of
events observed. At BES-III, with an integrated luminosity of 20
fb$^{-1}$ at $\psi(3770)$ peak, about $72\times10^6 D^0\bar D^0$
pairs will be collected with four year's
running~\cite{besiii,bepcii}. Table~\ref{tab} lists some promising
channels to search for $T$ asymmetry for both neutral and charged
$D$ decays and the corresponding expected statistical errors are
estimated. The projected efficiencies are extracted from
Ref.~\cite{besiii} and branching ratios are obtained from
Ref.~\cite{pdg}.
\begin{widetext}
\begin{center}
\begin{table}[H]
\begin{center}
\begin{tabular}{cccc}\hline\hline
$VV$             &Br (\%) & Eff. ($\epsilon$) &Expected errors\\
\hline
$\rho^0 \rho^0\rightarrow (\pi^+\pi^-)(\pi^+\pi^-)$  &0.18   & 0.74 &0.004\\
$\bar{K}^{*0} \rho^0 \to (K^-\pi^+) (\pi^+\pi^-)$    &1.08   &0.68  &0.002  \\
$\rho^0 \phi\rightarrow (\pi^+\pi^-)(K^+K^-)$    &0.14   & 0.26 &0.006\\
$\rho^+ \rho^- \rightarrow (\pi^+\pi^0)(\pi^-\pi^0)$  & 0.6$^*$    &0.55 &0.002\\
$K^{*+} K^{*-}\to (K^+ \pi^0)(K^- \pi^0)$    & 0.08$^*$ & 0.55 &0.006     \\
$K^{*0} \bar{K}^{*0}\to (K^+ \pi^-)(K^- \pi^+)$   &0.048  &0.62 &0.002    \\
 \hline
$\bar{K}^{*0}\rho^+\to (K^- \pi^+)(\pi^+\pi^0)$ &1.33 &0.59   &0.001 \\
\hline\hline
\end{tabular}
\caption{The promising (VV) modes with large branching fractions,
efficiencies and expected errors on the $T$ asymmetry: the
corresponding expected errors are estimated by assuming 20
fb$^{-1}$ data at $\psi(3770)$ peak at BES-III; the branching
fractions with asterisk are estimated according to
Refs.~\cite{Uppal:1992se,Kamal:1991,Das,Hin}. The last row is from
$D^+$ decay.}\label{tab}
\end{center}
\end{table}
\end{center}
\end{widetext}

In table~\ref{tab}, the branching fractions with asterisks have
not been measured yet, but some estimates combining naive
factorization and models for FSI are available from
Refs.~\cite{Uppal:1992se,Kamal:1991,Das,Hin}. Note that in
table~\ref{tab}, the estimated efficiencies are average value for
the various partial waves by assuming that the magnitude of the
longitudinal polarization is half of the decay rate. In the
future, a careful measurements at BES-III about the efficiency and
fraction for each partial wave are suggested. A more realistic
analysis requires a likelihood fit to the full angular dependence
of the $D\to V_1V_2\to (P_1P_2) (P_3P_4)$ mode. Systematics will
arise from the mis-reconstruction as $V_1 V_2$ of the events that
actually come from other resonances or non-resonance $D\to
P_1P_2P_3P_4$ background contributions. In view of the sizable
width of the vector resonances, we
 expect that these systematics will dominate the final result.
Their precise estimate in the BES-III experiment is  beyond the
scope of this paper. However, as pointed out in Ref.~\cite{Bigi2}
by Bigi, the four body decay of $D\to P_1P_2P_3P_4$ both with and
without intermediate states can all be used to probe $T$ asymmetry
in the frame work of TP.

 In conclusion, we studied
the $CP$ violation in $D\to VV$ decay mode in which the T
violating triple-product correlation is examined. That would
undoubtedly be another excellent probe of new physics beyond the
SM. The $CP$ violating observables in connection with angular
distribution are constructed. For neutral $D$ decays, we neglect
the $CP$ violation induced by $D^0-\bar D^0$ oscillation.
Experimentally, by doing a full angular analysis one may obtain
such $CP$ violating signals, and particularly it is worth
mentioning that the upcoming large $D$ data sample at BES-III will
provide a great opportunity to perform it. The sensitivities for
$CP$ violating observables are estimated by assuming 20$^{-1}$ fb
data-taking at $\psi(3770)$ peak at BES-III.

One of authors, X.~W.~Kang, wish to thank Rong-Gang Ping for the
help of drawing tools and Professor Z.~z.~Xing for many valuable
suggestions, and specially thank Professor G. Valencia for
stimulating discussions through mail. This work is supported in part
by the National Natural Science Foundation of China under contracts
Nos. 10521003, 10821063, 10835001, 10979008, the National Key Basic
Research Program (973 by MOST) under Contract No. 2009CB825200,
Knowledge Innovation Key Project of Chinese Academy of Sciences
under Contract No. KJCX2-YW-N29, the 100 Talents program of CAS, and
the Knowledge Innovation Project of CAS under contract Nos. U-612
and U-530 (IHEP).

\end{document}